\documentclass[conference]{IEEEtran}
\ifCLASSINFOpdf
\else
\fi

\usepackage{cite}
\usepackage{amsmath,amssymb,amsfonts}
\usepackage{caption}
\usepackage{subcaption}
\usepackage{graphicx}
\usepackage{textcomp}
\usepackage{xcolor}
\usepackage{amsmath}
\usepackage{multirow}
\usepackage{booktabs}
\usepackage{amsthm}
\usepackage{stfloats}
\usepackage{booktabs}
\usepackage{algorithm}
\usepackage{algorithmic}
\usepackage{amsmath}
\usepackage{url}

\begin{document}
%

\title{Flexible Design on Deterministic IP Networking \\ for Mixed Traffic Transmission}

 \author{\IEEEauthorblockN{Binwei Wu\IEEEauthorrefmark{1},
 Shuo Wang\IEEEauthorrefmark{2}\IEEEauthorrefmark{1},
 Jiasen Wang\IEEEauthorrefmark{1},
 Weiqian Tan\IEEEauthorrefmark{3}\IEEEauthorrefmark{1},
 Yunjie Liu\IEEEauthorrefmark{2}\IEEEauthorrefmark{1}}

 \IEEEauthorblockA{\IEEEauthorrefmark{1} Purple Mountain Laboratories, Nanjing, China\\}

 \IEEEauthorblockA{\IEEEauthorrefmark{2} Beijing University of Posts and Telecommunications, Beijing, China\\}

 \IEEEauthorblockA{\IEEEauthorrefmark{3} Southeast University, Nanjing, China\\}}

\maketitle

\begin{abstract}
Deterministic IP (DIP) networking is a promising technique that can provide delay-bounded transmission in large-scale networks.
Nevertheless, DIP faces several challenges in the mixed traffic scenarios, including (i) the capability of ultra-low latency communications, (ii) the simultaneous satisfaction of diverse QoS requirements, and (iii) the network efficiency. The problems are more formidable in the dynamic surroundings without prior knowledge of traffic demands.
To address the above-mentioned issues, this paper designs a flexible DIP (FDIP) network. In the proposed network, we classify the queues at the output port into multiple groups. Each group operates with different cycle lengths.
FDIP can assign the time-sensitive flows with different groups, hence delivering diverse QoS requirements, simultaneously. The ultra-low latency communication can be achieved by specific groups with short cycle lengths. Moreover, the flexible scheduling with diverse cycle lengths improves resource utilization, hence increasing the throughput (i.e., the number of acceptable time-sensitive flows).
We formulate a throughput maximization problem that jointly considers the admission control, transmission path selection, and cycle length assignment. A branch and bound (BnB)-based heuristic is developed.
Simulation results show that the proposed FDIP significantly outperforms the standard DIP in terms of both the throughput and the latency guarantees.

\end{abstract}


%
\IEEEpeerreviewmaketitle

\section{Introduction}
5G and mobile edge computing (MEC) networks pave the way for real-time network services, which enables a wide range of time-sensitive applications, such as mobile gaming, smart grid, remote surgery, and factory automation\cite{nasrallah2018ultra}. Most of these applications require just-in-time delivery of data traffic. Nevertheless, the traditional IP services for packet switch networks exhibit their limitations since they can only provide a best-effort transmission with no QoS guarantees. Some works may tempt to think that lightly-loaded communication links in the networks will yield low service latency and small end-to-end jitter. This is unfortunately not the case. The trivial experiments have shown that significant latency and jitter could be experienced even in a very lightly-loaded network due to the ``microburst''. Therefore, Internet Engineering Task Force (IETF) has developed a data-plane mechanism, known as deterministic IP networking (DIP), that can provide bounds on latency, packet delay variation (jitter), and high reliability in large-scale networks. The experiments have shown that the DIP can provide a delay-bounded transmission service over a transmission distance of more than 2,000 kilometers.

Nevertheless, the standard DIP exhibits several issues in the mixed traffic scenarios, e.g., MEC networks. First, the capability of supporting ultra-low latency communications waits to be further investigated. The QoS of DIP transmission strictly rests with the cycle length $\Delta_{\rm{dip}}$, i.e., the maximum end-to-end delays and jitter admit $\mathcal{O}(\Delta_{\rm{dip}})$ and
$2\Delta_{\rm{dip}}$, respectively. Note that $\Delta_{\rm{dip}}$ is lower bounded by $L_{\rm{pkt}}^{\max}/{\rm{BW}_e}$\footnote{$L_{\rm{pkt}}^{\max}$ is the maximum packet size in the network, and $\rm{BW}_e$ is the link bandwidth.}, hence limiting the deployment of DIP in the ultra-low latency scenarios.
Second, the standard DIP can only provide a specific QoS level. However, in mixed traffic scenarios, there is a wide range of time-sensitive applications with vastly different QoS requirements~\cite{grossman2018deterministic}. For instance, some industrial control applications may have very tight delay bounds (e.g., only a few microseconds) while the others may have more relaxed delay bounds up to a millisecond.
Third, it is challenging to find an efficient value of $\Delta_{\rm{dip}}$. In a DIP network, a long cycle length yields dissatisfaction with the ultra-low latency requirement while a short cycle length usually leads to low utilization due to resource fragmentation. The dynamic network surroundings without prior knowledge of the traffic demands (or profiles) exacerbate the problem.


To solve the above-mentioned issues, this paper proposes a flexible DIP network, where the queues at the port are classified into multiple groups. Each group operates with a specific cycle length. 
The time-sensitive flows are adaptively served by different groups to acquire different QoS guarantees.
The benefits of the proposed FDIP are summarized as follows. \\
(i) \textit{Achievement of the ultra-low latency transmission}:
For the ultra-low latency communications, FDIP only requires $\Delta_{1} \geq L_{\rm{pkt}}^{\rm{ull}}/{\rm{BW}_e}$, where $\Delta_{1}$ is the shortest cycle length in the FDIP, and $L_{\rm{pkt}}^{\rm{ull}}$ is the packet size of the flows with ultra-low latency requirement. Usually, $L_{\rm{pkt}}^{\rm{ull}} \ll L_{\rm{pkt}}^{\max}$. Hence,  ${\rm{inf}}\{\Delta_{1}\} \leq {\rm{inf}}\{\Delta_{\rm{dip}}\}$, which implies FDIP can perform better on the low-latency scenarios, compared to DIP. \\
(ii) \textit{Satisfaction of the diverse QoS requirements}:
FDIP can simultaneously achieve diverse QoS requirements by forwarding the flows with different cycle lengths. The short DIP cycle lengths could be used for the communications with the stringent low-latency requirements, whereas the long DIP cycle lengths are suitable for broadband services with some coarse-grain QoS constraints;\\
(iii) \textit{Improvement on the network throughput}: FDIP exploits the flexible scheduling on the cycle lengths to improve resource utilization and throughput. In FDIP, the time cycles with different lengths share the same physical resources. The fragmentation of the short time cycles can be reused by the flows assigned with the long time cycles.

\section{Background}

\subsection{DIP network}
DIP offers the end-to-end deterministic transmission by a cyclic hop-by-hop forwarding mechanism. Nodes in the DIP network divide the time into cycles with the length $\Delta_{\rm{dip}}$. Due to propagation delay, the downstream Node B may receive the packets from Node A at two different cycles (e.g., cycle $y-1$ and $y$). To absorb this variation, cycle mapping is introduced in DIP. Given the cycle mapping (i.e., $x \to y$), packets from Node A at cycle $x$ should be sent out by Node B at cycle $y+1$. Similarly, the packets at cycle $x+1$ on Node A should be sent out by Node B at cycle $y+2$.  DIP guarantees bounded delay and jitter as all packets experience the same maximum forwarding time, which is deterministic and known in advance by statistically measuring the worst-case delay.

To implement the cycle mapping and cyclic forwarding, DIP defines that, out of $N$ queues, $N_{\rm{dn}}$ queues (typically 4 queues for most cases) at each port are reserved for the time-sensitive flows. Each queue corresponds to a cycle. During the transmission, these $N_{\rm{dn}}$ queues are served in a round-robin fashion such that the active queue is open for transmission and closed for reception. Conversely, the $N_{\rm{dn}} - 1$ inactive queues can only accept packets for future transmission. The detail of DIP can be referred to \cite{QiangSDF} and \cite{LiQIANG:12}.

\subsection{Related works}

Extensional studies have been done to deliver the deterministic services in mixed traffic conditions.  Most researchers consider the system with two traffic classes, one for high-priority traffic with bounded delays, zero congestion packet loss, and small jitters (i.e., time-sensitive flows), and another one for low-priority best-effort traffic (i.e., best-effort flows). For example, IEEE 802.1Qbv standard \cite{oliver2018ieee} uses time gates at a switch egress that open/close according to a prescribed schedule, allowing zero interference among the two traffic classes. The authors of \cite{8681083} use the adaptive bandwidth sharing and adaptive slot windows, which permit the best-effort flow to occupy the unused bandwidth of the time-sensitive flows. The authors of \cite{joung2019regulating} consider a conserving fair scheduler knows as the regulating schedules, which acts as both as a regulator and a schedular to achieve fairness in the scheduling.

The dualistic classification of traffic types is not viable due to the vast QoS requirements of the further applications~\cite{info12010012}. Most studies focus on providing diverse QoS requirements in small-scale environments. The authors of \cite{9142734} deploy different real-time schedulers that are embedded in a software-defined network (SDN) controller to allow different applications with various QoS requirements. The authors of \cite{9229123} develop a general flexible window-based GCLs in the TSN networks to support mixed-critical messages. The authors of \cite{9318350} exploit the TSN mechanisms in the vehicle networks with six different QoS configurations. For the large-scale networks, the authors of \cite{nasrallah2018ultra} believe that it is possible to develop a registration and reservation protocol that can reconfigure the network by scaling up/down the cycle length to accommodate the dynamical surroundings. However, how to simultaneously and efficiently support multiple QoS rules in a single large-scale deterministic network is still an open question.

\section{DIP network with flexible cycle length}

\subsection{Design}

We use $\mathcal{G} = \left\{\mathcal{V}, \mathcal{E}\right\}$ to denote the network, where $\mathcal{V}$ and $\mathcal{E}$ are nodes and links, respectively. The time in the network is divided into unitary cycles with duration $\Delta_{0}$. A link $e = (v_i, v_j)\in\mathcal{E}$ incurs a link delay $\tau_{v_i, v_j}$ and its bandwidth is ${\rm{BW}}_{e}^{\rm{link}}$, measured by bits per second. We neglect the processing delay on the nodes for simplicity.

This paper only considers the deterministic periodic time-sensitive flows, i.e., the flows with known source, sink, arrival time, packet size, cyclic time, and requirements on end-to-end latency (and jitter). 
We describe the time-sensitive flows as demands $\mathcal{D}$. A demand $d\in\mathcal{D}$ is defined by a 7-tuple $\left<s^d, t^d, T^d, c^d, \omega^d, \Gamma^d, \Pi^d \right>$, where $s^d$ and $t^d$ are the source and sink, $T^d$ is its cyclic time (in unitary cycles), $c^d$ is the arrival time (i.e.,  index of the unitary cycles, starting at 0), $\omega^d$ is the payload (in bits), $\Gamma^d$ is the maximum tolerable end-to-end latency, and $\Pi^d$ is the maximum acceptable jitter. $t$ can be the edge clouds in the MEC networks or application servers in the data centers.
We use $\mathcal{P}_d$ to denote the paths from $s^d$ to $t^d$. A path $p = (v_0, v_1, v_2, \cdots, v_{|p|})\in\mathcal{P}_d$ is a set, consisting of nodes along the path. We have $v_0 = s^d$ and $v_{|p|} = t^d$.

\begin{figure}[ht]
\centering
\includegraphics[width=.70\linewidth]{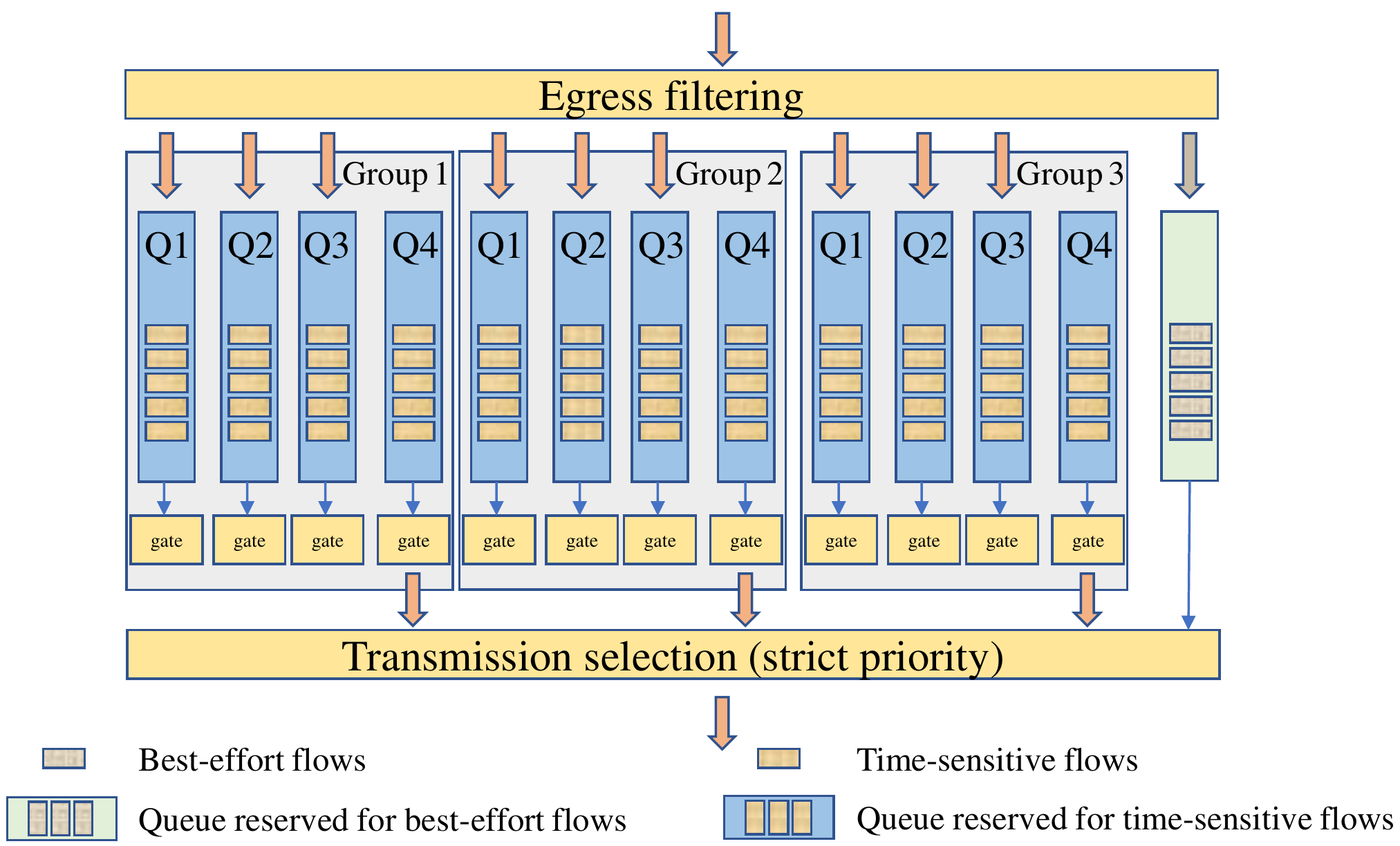}
\caption{\small  FDIP classifies the queues into different groups, each of which operates with different cycle lengths. In the figure, 12 queues (divided into 3 groups, $M=3$) are reserved for the time-sensitive flows and 1 queue is reserved for the best-effort flows. Modules, including the egress filtering, gate, and transmission selection, are defined in IEEE 8021Qch and instantiated in DIP.}
\label{fig:fig2}
\end{figure}
\begin{figure}[ht]
\centering
\includegraphics[width=.75\linewidth]{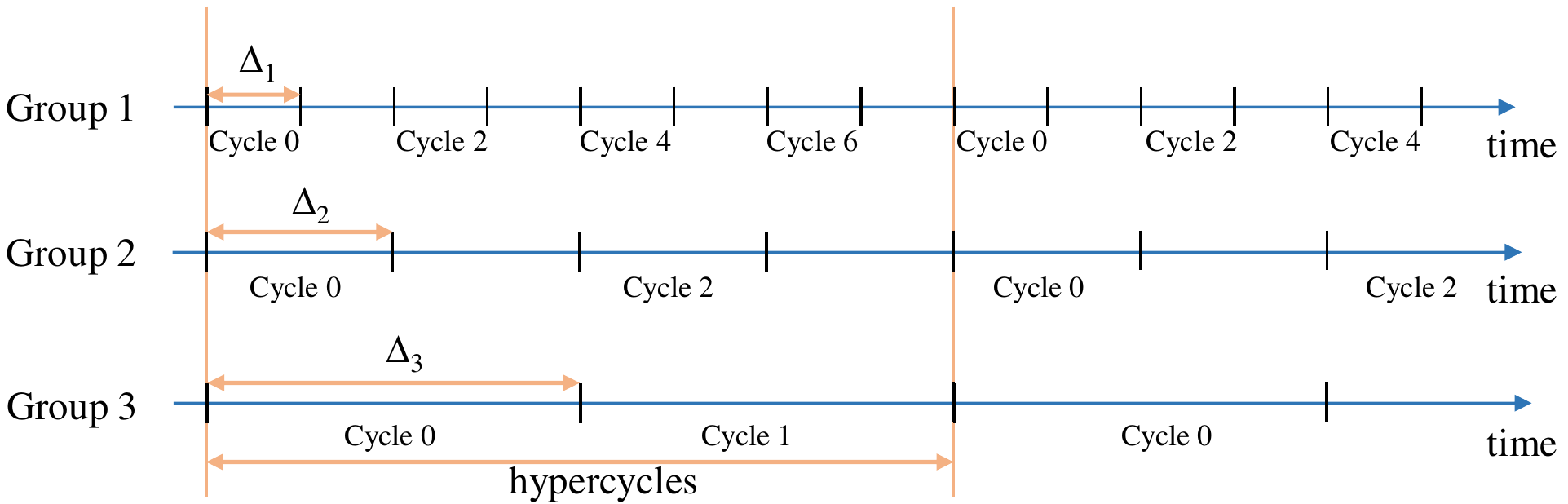}
\caption{\small  The groups divide the time into time cycles with different lengths. The groups at the same port achieve the perfect time synchronization on the level of the hypercycle.}
\label{fig:fig3}
\end{figure}
All the nodes in the FDIP must support the deterministic mechanisms in DIP, e.g., time division, cycle mapping, and cyclic forwarding.
Additionally, assume that each port in FDIP is equipped with $N$ queues. Then, out of $N$ queues, $MN_{\rm{dn}}$ ($N_{\rm{dn}} \geq 3$) queues are reserved for the time-sensitive flows. As shown in Fig.\ref{fig:fig2}, these queues are partitioned into $M$ groups, each of which consists of $N_{\rm{dn}}$ queues.
In every group, the queues operate like the ones in DIP with cycle length ${\Delta _m}$, i.e., (i) the time in group $m$ is divided into cycles with duration ${\Delta _m}$; (ii) the queues in group $m$ open and close alternatively in a cyclic fashion; (iii) each queue in group $m$ corresponds to a cycle with length ${\Delta _m}$; (iv) only 1 queue is active for transmission while the other ones are close for reception. The cycle length ${\Delta _m}$ is defined by
\begin{equation}
{\Delta _m} = {k_m}{\Delta _{m - 1}},\;m \in \mathcal{M} = \{1,\cdots,M\},
\end{equation}
where $k_m\in\mathbb{Z}_{+}$.

Inside a FDIP-enabled device, the queues of the $M$ groups achieve a strict time synchronization on the level of the hypercycle, i.e., the start time of the hypercycles is the same, as shown in Fig.\ref{fig:fig3}. The length of the hypercycle is given by
\begin{equation}
{\Delta _{\rm{hc}}} = N_{\rm{hc}}\; {\rm{lcm}}\left(\{ {\Delta _M}\} \cup \{T^d, \forall d\in\mathcal{D}\}\right),
\end{equation}
where $N_{\rm{hc}}\in\mathbb{Z}_{+}$ is sufficiently large so that for every $m\in\mathcal{M}$, ${\Delta _{\rm{hc}}}/{\Delta _{m}} \geq N_{\rm{dn}}$. For simplicity, let $T^d = \Delta_{\rm{hc}}, \forall d\in\mathcal{D}$.
We also define a cycle alignment function from group $m_i$ to $m_j$ ($m_i < m_j$),  given by
\begin{equation}
\varphi_{m_i, m_j}(a) = \lceil \frac{(a+1)\Delta_{m_i}}{\Delta_{m_j}}  \rceil -1, \quad a\in\mathcal{N}_{{m_i}},
\end{equation}
where $a$ is a cycle index of group $m_i$ ($a \in [0, |\mathcal{N}_{m_i}|-1]$), and $\mathcal{N}_{m_i}$ is the set of cycles of group $m_i$  in a hypercycle. $\varphi_{m_i, m_j}(a)=b$ represents that cycle $a$ in  group $m_i$ and cycle $b$ in group $m_j$ overlaps in time. Meanwhile, we use $\mathcal{C} = \varphi^{-1}_{m_i, m_j}(b)$ to denote the set of cycles in group $m_i$ that share the same time periods with cycle $b$ in group $m_j$.

At any given time, only $M$ queues are active at an output port for the packet transmission while the other ones are for the reception. The FDIP-enabled devices use the strict priority policy for transmission selection among the active queues. FDIP believes that the priority of queues in the group $m-1$ is higher than the ones in the group $m$. The output port transmits a packet from an active queue if: (i) the queue contains a packet ready for the transmission, (ii) the active queue with higher priority does not have a frame to transmit, and (iii) the transmission can compete before the queue is closed. Preemption is used to improve utilization. Preemptable packets that are in transit can be preempted by express packets. After the transmission of express packets has been completed, the transmission of the preempted packet can resume.

\subsection{Cyclic forwarding in the FDIP}

FDIP provides the deterministic QoS through cyclic forwarding with different cycle lengths. During the forwarding, every time-sensitive flow is adaptively assigned with a group $m\in\mathcal{M}$ (by using the egress filtering in Fig.\ref{fig:fig2}\footnote{Each packet would carry a SR label (at least $\log_2{M}$ bits) to specify the group. The egress filtering would read the label and forward the packets to the corresponding group.}).

To enable cyclic forwarding, FDIP exploits frequency synchronization among the neighboring nodes; and uses the cycle mapping mechanism, as follows.

{\textit{Frequency synchronization}}: Consider the group $m$ on two neighboring nodes, $v_i$ and $v_j$. FDIP requires that the group $m$ on node $v_i$ achieves the frequency synchronization of $1/\Delta_{\rm{m}}$ with the group $m$ on node $v_j$, i.e., the offset of the start time of the time cycles of group $m$ on node $v_i$ and $v_j$ remains constant.

{\textit{Cycle mapping}:} Consider the deterministic transmission that assigned with group $m$ between the neighboring node $v_i$ and $v_j$. We define the cycle mapping, $\Phi_{v_i, v_j}^{m}(\cdot)$, given by
\begin{equation}
\Phi_{v_i, v_j}^{m}(a) = {\rm{mod}}\left(\lfloor \frac {(a+1)\Delta_{m}+\tau_{(v_i,v_j)}-\tau^{\rm{hc}}_{v_i,v_j}}{\Delta_{m}}\rfloor, N_{m}\right),
\end{equation}
where ${\rm{mod}}(\cdot)$ is the residue arithmetic operation; and $\tau_{v_i, v_j}^{\rm{hc}}\in (-\Delta_{m}, \Delta_{m})$ is the offset of hypercycles on node $v_i$ and $v_j$; and, $N_m = |\mathcal{N}_m|$. $\Phi_{v_i, v_j}^{m}(a) = b$ represents that the packets (assigned with group $m$) sent out at cycle $a$ on node $v_i$ is re-sent out at cycle $b+1$ on node $v_j$.
$\Phi_{v_i, v_j}^{m}(\cdot)$ is a periodic function with duration  $\Delta_{\rm{hc}}$.
We define a function to output the delay on hop $v_i{\to}v_j$ (for packets assigned with $m$), denoted by $\phi_{v_i,v_j}^m$:
\begin{equation}\label{phi}
\begin{aligned}
\phi_{v_i,v_j}^m(a) = & \lfloor \frac {(a+1)\Delta_{m}+\tau_{(v_i,v_j)}-\tau^{\rm{hc}}_{v_i,v_j}}{\Delta_{m}}\rfloor
 \Delta_{m} \\
& + \tau^{\rm{hc}}_{v_i,v_j} - a \Delta_{m},
\end{aligned}
\end{equation}
where $a$ is the transmitted cycle index in group $m$ on node $v_i$. $\phi_{v_i,v_j}^m(a)$ outputs the durations from the end of cycle $a$ on node $v_i$ to the time of packets to be transmitted (the end time of cycle $\Phi_{v_i,v_j}^m(a)$) on node $v_j$.

\section{Problem formulation}
\subsection{Decision variables}
The following decision-making variables are jointly considered:

{\textit{Admission control}:} We use $x^d$ to describe whether demand $d$ is accepted ($x^d=1$) or not ($x^d = 0$). Let $\mathcal{X}=\{x^d, \forall d\in\mathcal{D}\}$.

{\textit{Path selection}:} A path $p^d = (v_0,\cdots,v_{|p^d|}) \in \mathcal{P}_d$ is decided for every accepted demand $d$. Let $\mathcal{S} = \{p^d, \forall d\in\mathcal{D}\}$.

{\textit{Cyclic length (group) assignment}:} We use $m^d\in\mathcal{M}$ to describe the cycle length that is assigned to demand $d$. Let $\mathcal{Y} = \{m^d, \forall d\in\mathcal{D}\}$.

\subsection{Constraints on the latency and jitter}
%
%

We present the QoS metric of an accepted demand $d$ as $\Delta^d$ (the end-to-end delay) and $J^d$ (the maximum jitter). Suppose demand $d$ is accepted and assigned with path $p^d = \{v_0, \cdots, v_{|p^d|}\}$ and group $m^d$. Then, $\Delta^d \leq \Gamma^d$ and $ J^d \leq \Pi^d$.

In every realization of the hypercycle, the intermediate node $v_i$ on the path $p^d$ re-sends the demand $d$'s packets out at cycle $c_{i}^d$ (in group $m^d$), given by
\begin{equation}\label{e1}
c_{i}^d = {\rm{mod}}\left(\Phi_{v_{i-1}, v_i}^{m^d}(c_{i-1}^d)+1, N_{m^d}\right),
\end{equation}
where $c^d_{0} = \varphi_{m_0, m^d}(c^d)+1$ and $ i\in\{1,\cdots,|p^d|-1\}$. Let $\Delta_{i}^d$ be the (maximum) accumulated delay for demand $d$ to be transmitted on intermediate node $v_i$. It is calculated as
\begin{equation}\label{e1}
{\Delta _{i} ^d} = {\Delta _{i-1} ^d} + \Delta_{m^d} + \phi_{v_{i-1},v_i}^{m^d}(c_{i-1}^d)
\end{equation}
where ${\Delta _{0} ^d} = \Delta_{m^d}$, and $i\in\{1,\cdots,|p^d|\}$. Then, the constraints on the end-to-end latency for demand $d$ is
\begin{equation}\label{c1}
\Delta^d \leq {\Delta _{|p^d|} ^d} \leq \Gamma^{d}, \quad \forall d\in\mathcal{D}.
\end{equation}

The constraints on jitters are inherently given by
\begin{equation}\label{c2}
2 \Delta_{m^d} \leq J^d, \quad \forall d \in \mathcal{D}
\end{equation}

\subsection{Constraints on capacity}

For every realization of the hypercycle, the accepted demand $d$ consumes a certain capacity $w_{e, m}^d(c)$ on wired link $e=(v_{i-1}, v_{i})$ along its s-path $p^d$, where $c$ is the index of the cycles with length $\Delta_{m}$. Here, $w_{e, m}^d(c)$ is given by
\begin{equation}
w_{e, m}^d\left( c \right) = \left\{ {\begin{array}{*{20}{l}}
{{\omega ^d}  }&{if\;(m = m^d) \& (c = c_{i - 1}^d)}\\
0&{otherwise}.
\end{array}} \right.
\end{equation}
The aggregated traffic at cycle $c$ in group $m$ on link $e$ is
\begin{equation}
w_{e, m}(c) = \sum\limits_{d\in\mathcal{D}} x^d w_{e, m}^d\left( c \right) + \sum\limits_{\tau\in\varphi_{m-1, m}^{-1}(c)} w_{e, m-1}\left( \tau \right),
\end{equation}
where $w_{e, 0}(c) = 0$. The last term is introduced due to the strict priority policy among the groups. Finally, the constraints on the link bandwidth can be given by
\begin{equation}\label{c3}
\begin{aligned}
w_{e,m}(c) \leq  {\rm{BW}}^{\rm{link}}_e \cdot \Delta_{m}, \forall e \in \mathcal{E}, c \in \mathcal{N}_{m}, m\in\mathcal{M}.
\end{aligned}
\end{equation}

\subsection{Objective function}

The goal of the controller is to accept a subset of demands $\mathcal{D}$ such that the total number of the accepted demands is maximized. The problem of interest is given by:

\begin{subequations}\label{P1}
\begin{alignat}{2}
\mathcal{W}_{\rm{opt}} = \max\limits_{{\mathcal X}, {\mathcal{Y}}, {\mathcal{S}}} & \sum_{d\in\mathcal{D}} x^d\\
\mbox{s.t.} \quad  & x^d \in \{0, 1\}, \quad \forall d\in\mathcal{D} \label{c6}\\
 & m^d \in \mathcal{M}, \quad \forall d\in\mathcal{D} \label{c7} \\
 & |p^d| \leq H, \quad \forall d\in\mathcal{D} \label{c8} \\
 & \eqref{c1},\eqref{c2},\eqref{c3}. \notag
\end{alignat}
\end{subequations}
Constraint \eqref{c6} gives the self-explanatory of the admission control. Constraint \eqref{c7} ensures that the number of hops along the s-path is no greater than $H$ per path.

{\textit {NP-hardness}:}  Problem~\eqref{P1} is an NP-complete problem. The network needs to decide if, for a given threshold $l \in  \mathbb{R}_+$, there is a feasible solution to Problem~\eqref{P1} with the objective value $\mathcal{W}_{\rm{opt}}\leq l$. The following reduction proof is based on the well-known $k$-Disjoint Paths (k-DP) problem~\cite{korte2009combinatorial}. We consider the (NP-complete) version of $k$-DP which decides if $k$ link-disjoint paths can be found between nodes $s$ and $t$ in a directed graph $\mathcal{G}$. This problem can be reduced to an instance of Problem~\eqref{P1} by setting $M = 1$, $\Delta_{\rm{hc}} = \Delta_{1} = T^d = 1$, and $\omega^d = 1, \forall  d\in\mathcal{D}$ that all have sources and destinations. The capacity of every link $e$ is chosen to be one. Choosing $k = l$, Problem~\eqref{P1} returns true if and only if there are $k$ arc-disjoint paths in the network $\mathcal{G}$. Since all reduction steps are polynomial in the problem size, the NP-hardness proof is complete. Furthermore, it is clear that  Problem~\eqref{P1} belongs to NP since the validity of any solution can be checked in polynomial time. Thus, Problem~\eqref{P1} is NP-complete.

\section{Proposed BnB-based solution}

The BnB approach is an important algorithm proposed in the literature to solve the combinatorial optimization problem. In general, a standard BnB algorithm comprise two steps: branching and bounding. The breaching repeatedly divides the solution spaces into smaller subsets. Then, the bounding is conducted, which obtains the upper bound for each subset, i.e., solving a lower-dimensional subproblem. After each breaching, some branches are efficiently removed from the search tree. The premise of pruning are as follows:
\begin{enumerate}
\item The subproblem after branching is infeasible.
\item The upper bound of the subproblem after breaching is smaller than the best-known objective value.
\end{enumerate}

The proposed BnB-based algorithm is presented in Algorithm~1. Let set $\mathcal{A}$ denote all the decision-making variables instead of $x^d$, $p^d$, and $m^d$. An element of $\mathcal{A}$, $a_{d, p, m}\in \mathcal{A}$, is a binary variable, i.e., $a_{d, p, m} = 0$ represents that demand $d$ is not accepted; otherwise ($a_{d, p, m} = 1$), demand $d$ is assigned with path $p$ and cycle length $m$. Specially, we define the set of solution space as $\mathcal{L} = \{(d,p,m)| \forall d\in\mathcal{D}, p\in\mathcal{P}, m\in\mathcal{M}\}$ as the branch and bound nodes. In addition, we define two sets $\mathcal{L}_0 = \{(d,p,m)| a_{d, p, m} = 0, \forall d\in\mathcal{D}, p\in\mathcal{P}, m\in\mathcal{M}\}$ and $\mathcal{L}_1 = \{(d,p,m)| a_{d, p, m} = 1, \forall d\in\mathcal{D}, p\in\mathcal{P}, m\in\mathcal{M}\}$. With $\mathcal{L}_0$ and $\mathcal{L}_1$, problem \eqref{P1} can be equivalent to
\begin{subequations}\label{P2}
\begin{alignat}{2}
\mathcal{W}_{\rm{opt}}& =\max\limits_{{\mathcal X}, {\mathcal{Y}}, {\mathcal{S}}} \sum_{d\in\mathcal{D}} a_{d, p, m}\\
\mbox{s.t.} \quad  & a_{d, p, m} = 0, \quad \forall (d,p,m) \in\mathcal{L}_0 \label{c9}\\
 & a_{d, p, m} = 1, \quad \forall (d,p,m) \in\mathcal{L}_1 \label{c10} \\
 & a_{d, p, m} = \{0, 1\}, \quad \forall (d,p,m) \in\mathcal{L}\setminus(\mathcal{L}_0 \cup \mathcal{L}_1) \label{c11} \\
 & \eqref{c1},\eqref{c2},\eqref{c3} \eqref{c6},\eqref{c7},\eqref{c8}. \notag
\end{alignat}
\end{subequations}
When $\mathcal{L}_1$ and $\mathcal{L}_0$ is fixed, the continuous relaxation of problem \eqref{P2} can be formulated as
\begin{subequations}\label{P3}
\begin{alignat}{2}
\mathcal{W}_{\rm{opt}}^{\rm{rel}}& = \max\limits_{{\mathcal X}, {\mathcal{Y}}, {\mathcal{S}}} \sum_{d\in\mathcal{D}} a_{d, p, m}\\
\mbox{s.t.} \quad  & a_{d, p, m} = 0, \quad \forall (d,p,m) \in\mathcal{L}_0 \label{c9}\\
 & a_{d, p, m} = 1, \quad \forall (d,p,m) \in\mathcal{L}_1 \label{c10} \\
 & a_{d, p, m} \in [0, 1], \quad \forall (d,p,m) \in\mathcal{L}\setminus(\mathcal{L}_0 \cup \mathcal{L}_1) \label{c11} \\
 & \eqref{c1},\eqref{c2},\eqref{c3} \eqref{c6},\eqref{c7},\eqref{c8}. \notag
\end{alignat}
\end{subequations}
Obviously, $\mathcal{W}_{\rm{opt}}$ in \eqref{P2} is upper bounded by $\mathcal{W}_{\rm{opt}}^{\rm{rel}}$ in \eqref{P3}. We respectively present problem \eqref{P2} and \eqref{P3} with tuples $(o, \mathcal{L}_0, \mathcal{L}_1)$ and $(o, \mathcal{L}_0, \mathcal{L}_1)^{\rm{rel}}$, where $o = \mathcal{W}_{\rm{opt}}^{\rm{rel}}$ is the optimal value of problem \eqref{P3}. Then, we define $\mathcal{I}$ as the set of the brunch problem and $o^*$ as the best-known objective value.
The main process of the BnB algorithm is as follows:

\addtolength{\topmargin}{0.03in}
{\textit{Branching}}: In each branching iteration, the problem that attains the maximum upper bound, denoted as $(\hat o, \hat {\mathcal{L}_0}, \hat {\mathcal{L}_1})$, is chosen to be branched. Then, we select the branching node $(d^*, p^*, m^*)$ with the highest priority. By setting the integer variable $a_{d,p,m}$ to 0 or 1, we divide the problem into two smaller problems, i.e., $(o^{(1)}, \mathcal{L}_0^{(1)}, \mathcal{L}_1^{(1)})^{\rm{rel}}$ and $(o^{(2)}, \mathcal{L}_0^{(2)}, \mathcal{L}_1^{(2)})^{\rm{rel}}$. It is apparent that the priority function has significant impacts on the complexity of the proposed algorithm. The priority function for $(\hat o, \hat {\mathcal{L}_0}, \hat {\mathcal{L}_1})$ is set as
\begin{equation}\label{weight}
\begin{aligned}
\xi(d,p,m) = \xi_1 v^d + \xi_2 \sum\limits_{e\in{p}}\left( 1 - \frac {w_{e,m}(c)}{{\rm{BW}}^{\rm{link}}_e\Delta_m}\right)^{\xi_3}
\end{aligned}
\end{equation}
where $\xi_1$, $\xi_2$ and $\xi_3$ are the weighting factors.

{\textit{Bounding and pruning}}:
Based on the selected branch, we calculate the maximum upper bound of the sub-problems, i.e., $(o^{(1)}, \mathcal{L}_0^{(1)}, \mathcal{L}_1^{(1)})^{\rm{rel}}$ and $(o^{(2)}, \mathcal{L}_0^{(2)}, \mathcal{L}_1^{(2)})^{\rm{rel}}$, respectively.
The global convergence is guaranteed by simple bound. Compare the new
solutions, i.e., $(o^{(1)}, o^{(2)})$, with the current best-known objective value $o^*$, and update the $o^*$ with the larger one. The problem whose upper bound is larger than $o^*$ will be put into $\mathcal{I}$.
Otherwise, it will be pruned.

\begin{algorithm}
  \caption{BnB-based resource allocation algorithm}
  \label{alg:alg1}
  \begin{algorithmic}[1]
 \STATE {\bf Initialization:} $\mathcal{I} = (o,\mathcal{L}_0, \mathcal{L}_1)$, $o^*=-\infty$, $\mathcal{L}_0 = \mathcal{L}_1 = \emptyset$.
 \WHILE {$\mathcal{I}\neq \emptyset$}
 \STATE Based on $\hat o = \min\nolimits_{(o,\mathcal{L}_0, \mathcal{L}_1)\in\mathcal{I}}o$, choose $(\hat o,\hat{\mathcal{L}}_0, \hat{\mathcal{L}}_1)$ and update $\mathcal{I} = \mathcal{I}\setminus(\hat o,\hat{\mathcal{L}}_0, \hat{\mathcal{L}}_1)$.
 \STATE Choose a branching node with the highest priority, i.e., $(d^*, p^*, m^*) = \arg \max\nolimits_{(d, p, m)\in \mathcal{L}\setminus(\mathcal{L}_0\cup\mathcal{L}_1)} \xi(d, m, p)$.
 \STATE Update $\mathcal{L}_0^{(1)} = \hat {\mathcal{L}}\cup\{(d^*, p^*, m^*)\}$,  $\mathcal{L}_1^{(1)} = \hat {\mathcal{L}}$.
 \STATE Update $\mathcal{L}_0^{(2)} = \hat {\mathcal{L}}$, $\mathcal{L}_1^{(2)} = \hat {\mathcal{L}}\cup\{(d^*, p^*, m^*)\}$.
 \STATE Solve the sub-problems with the interior point method, i.e., $(o^{(1)}, \mathcal{L}_0^{(1)}, \mathcal{L}_1^{(1)})^{\rm{rel}}$ and $(o^{(2)}, \mathcal{L}_0^{(2)}, \mathcal{L}_1^{(2)})^{\rm{rel}}$. If no feasible solution, set the result as $-\infty$.
 \IF {$o^{(i)}> o^*, i \in \{1, 2\}$}
 \IF {$\mathcal{L}==\mathcal{L}_0^{(i)}\cup \mathcal{L}_1^{(i)} $}
 \STATE Let $o^*=o^{(i)}$, $\mathcal{L}_0^*=\mathcal{L}_0^{(i)}$, and $\mathcal{L}_1^*=\mathcal{L}_1^{(i)}$.
 \ELSE
 \STATE Update $\mathcal{I} = \mathcal{I} \cup \{(o^{(i)},\mathcal{L}_0^{(i)}, \mathcal{L}_1^{(i)})\}$
 \ENDIF
 \ENDIF
 \STATE Prune the branches.
 \ENDWHILE
\STATE {\bf Return}: $o^*,\mathcal{L}_0^*$, and $ \mathcal{L}_1^*$
  \end{algorithmic}
\end{algorithm}

\section{Simulation}
\begin{figure}[ht]
\centering
\includegraphics[width=.55\linewidth]{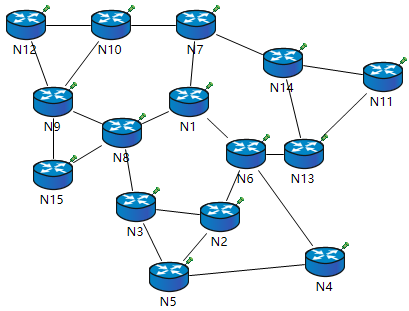}
\caption{\small Network topology produced from a real-work network in SNDlib, named as ``Atlanta''.}
\label{fig:topo}
\end{figure}

We evaluate the efficiency of the proposed FDIP network in the OMNet++. As shown in Fig.~\ref{fig:topo}, the simulation topology is produced with a real-world network from SNDlib~\cite{OrlowskiPioroTomaszewskiWessaely2010}. The link delays vary from 60~$\mu$s to 70~$\mu$s, randomly. We set the link capacity as 10 Gbps, uniformly. Both the time-sensitive flows and best-effort flows are involved in the simulations. The time-sensitive flows are pumped into the network from N2, N10, N14, and N15; and departure from N1. As depicted in Table~I, we mainly consider three (time-sensitive) traffic types, corresponding to the programmable logic controller (PLC) automation, industrial supervision, and remote virtual reality (VR)~\cite{nasrallah2018ultra,grossman2018deterministic}. Best effort flows (produced by UDP burst applications in the OMNet++) are injected into the links to create the ``microburst''. The factors $\xi_1 = \xi_2 = \xi_3 = 1$ in \eqref{weight}.

\begin{table}[]
\caption{Profiles for the time-sensitive flows.}
\begin{tabular}{@{}lccccc@{}}
\toprule
       & \begin{tabular}[c]{@{}c@{}}Cyclic\\ time\end{tabular} & Payload   & \begin{tabular}[c]{@{}c@{}}Maximum\\ latency\end{tabular} & \begin{tabular}[c]{@{}c@{}}Maximum\\ jitter\end{tabular} & N.O. flows \\ \midrule
Type 1 & 0.1ms       & 750 Bytes & 500$\mu$s                                                 & 100$\mu$s                                                & 2000       \\
Type 2 & 0.5ms       & 1.5KBytes & 900$\mu$s                                                 & --                                                       & 2000       \\
Type 3 & 1.0ms       & 6.2KBytes & 2ms                                                       & --                                                       & 2000       \\ \bottomrule
\end{tabular}
\end{table}

\begin{figure}[ht]
\centering
\includegraphics[width=.90\linewidth]{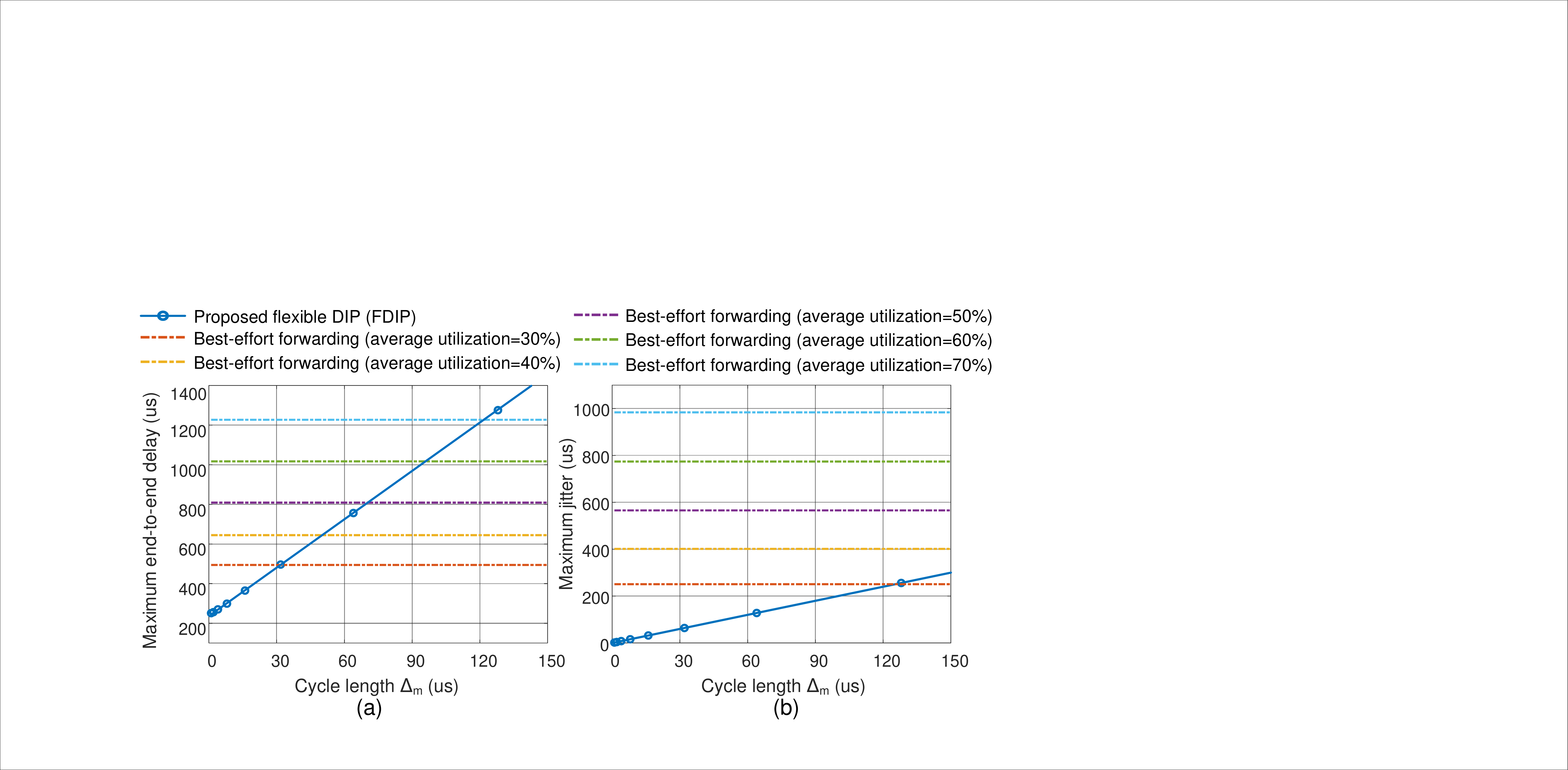}
\caption{\small Relationships between the cycle length and QoS metric. (a) cycle length v.s. maximum end-to-end latency; (b) cycle length v.s. maximum jitter.}
\label{fig:sim1}
\end{figure}

Fig.~\ref{fig:sim1} shows that the QoS of the time-sensitive flows in FDIP strictly rests with its assigned cycle length. We measure the end-to-end delay and maximum jitter with different cycle lengths $\Delta_{m}$ and link congestions, and compare the results with the traditional best-forwarding method. With the priority-based scheduling among the time-sensitive and BE flows, the congestions produced by BE flows have no effects on the time-sensitive transmission. It is also noticed that the maximum delay of the flows in FDIP is linear with their assigned $\Delta_{m}$. With a small $\Delta_{m}$ (e.g., $\Delta_{m} \leq 32{\mu}s$), an ultra-low latency communication can be achieved, e.g., the end-to-end delay is smaller than the ones with best-effort forwarding in a lightly-loaded scenario (link utilization = 30\%). When assigned with a large $\Delta_{m}$, FDIP can still provide a deterministic transmission with no packet loss and small jitter.

\begin{figure}[ht]
\centering
\includegraphics[width=.99\linewidth]{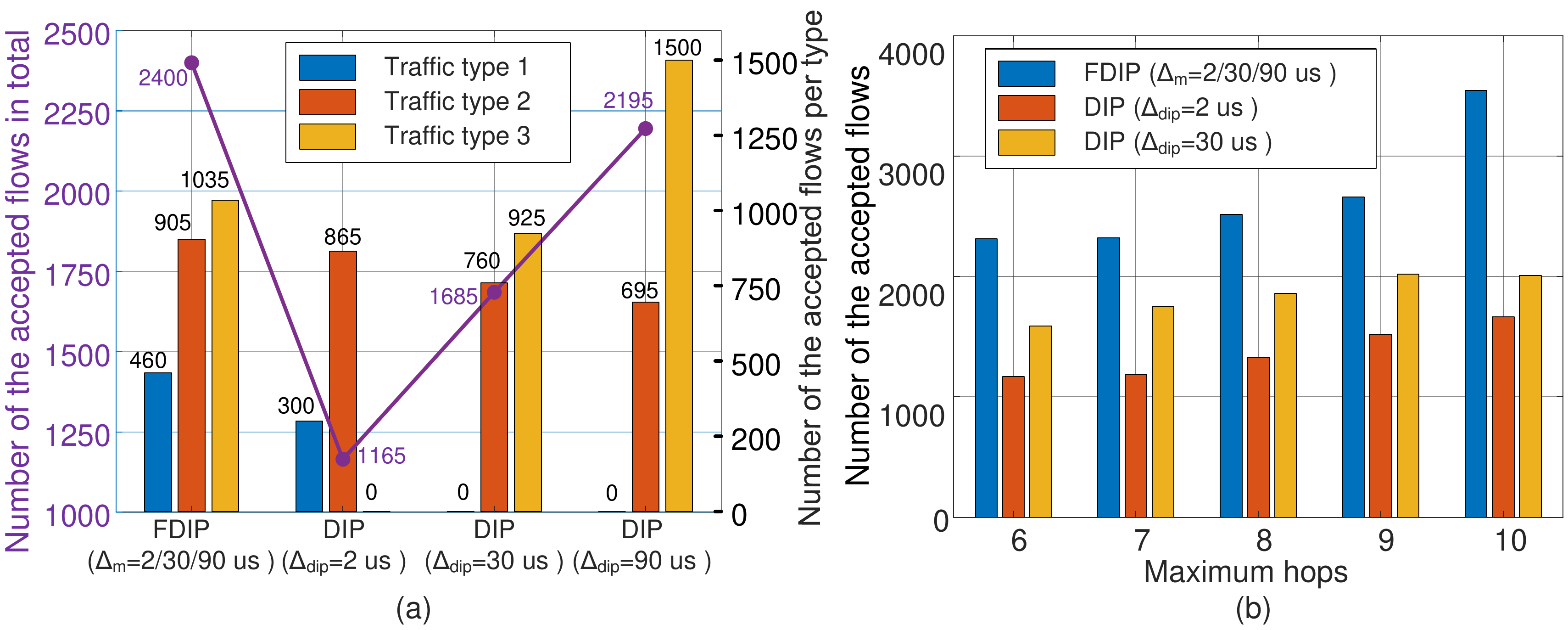}
\caption{\small Performance evaluation of FDIP. (a) Satisfaction of diverse QoS requirement. (b) Throughput comparison with standard DIP.}
\label{fig:sim2}
\end{figure}

Fig.~\ref{fig:sim2}(a) depicts that the FDIP can provide diverse QoS requirements. The FDIP can support all the QoS requirements in Table~1 while the standard DIP can only simultaneously serve two (out of three) traffic types. To achieve the ultra-low latency communication (e.g., Type 1 requires an end-to-end latency beneath $500 {\mu}$s), the standard DIP-enabled device configures the cycle length as $2\mu$s. Nevertheless, a small cycle length restricts the maximum packet length by $2{\mu}s{\times}10\rm{Gbps}=20$Kbits. Thus, the network is unable to serve the flows of Type 3. A possible solution is the segmentation, which brings new challenges (e.g., overheads) and increases the complexity.

Fig.~\ref{fig:sim2}(b) shows that FDIP dominates the standard DIP on the network throughput. Meanwhile, the advantage of FDIP enlarges with the increase of the maximum hops $H$. The advantages attribute to the flexible scheduling in the proposed FDIP. Typically, with a large $H$, the number of feasible paths increases. The traffic can be balanced in a larger area, which relieves the congestion and improves the throughput. However, for the paths with more hops involved, the end-to-end latency sustainingly increases. In the time-sensitive scenarios, the benefits of increment on $H$ are limited in the DIP. On the other hand, FDIP can exploit shorter cycle lengths for the longer path to reduce the path delay. Given the same QoS requirement, the number of feasible paths increases, hence improving the throughput. It is also noticed that FDIP can adaptively select the cycle length to enhance resource utilization. Compared to the standard DIP (with only one cycle length), FDIP can adaptively schedule with multiple cycle lengths to improve the utilization. Generally, a small cycle length leads to low resource utilization due to resource fragmentation. The proposed FDIP would use a large cycle length along the shorter paths, which enhances the utilization, hence improving the throughput.

\section{Conclusion}

This paper proposes a FDIP network to attain the diverse QoS requirement and improve network efficiency. Every FDIP-enabled devices classify the queues at the output port into multiple groups. Each group operates in a specialized cycle length.
By assigning the time-sensitive flows with different groups, the proposed FDIP can simultaneously support diverse QoS requirements, i.e., a short cycle length is exploited to support ultra-low latency communication, and vice versa. Meanwhile, the flexible scheduling on the cycle length contributes to improving the resource utilization, hence increasing the throughput.
We also formulate an integer programming to maximize the network throughput, which jointly considers the admission control, cycle length (group) assignment, and transmission path selection. A BnB-based heuristic approach is developed. Simulation results show that the proposed FDIP significantly outperforms the standard DIP network in terms of both the throughput and the QoS guarantees.



%

%

\end{document}